\PassOptionsToPackage{dvipsnames}{xcolor}
\documentclass[sigconf,nonacm,screen]{acmart}

\usepackage[all]{nowidow}
\usepackage{xcolor}
\usepackage{amsmath}
\usepackage{bm}
\usepackage{subcaption}
\usepackage{hyperref}
\usepackage{acronym}
\usepackage{tabularx}   
\usepackage{booktabs}   
\usepackage{pifont}
\newcommand{\cmark}{\ding{51}}
\newcommand{\xmark}{\ding{55}}
\usepackage{array}
\usepackage{csquotes}
\usepackage{siunitx}
\DeclareSIUnit{\millicore}{mc}
\newcommand{\tRange}[3]{\qtyrange[range-units = single, range-phrase = --]{#1}{#2}{#3}}
\newcommand{\tFactor}[1]{\num{#1}\,$\times$}

\usepackage[capitalise,noabbrev]{cleveref}
\crefformat{section}{\S#2#1#3} %
\crefformat{subsection}{\S#2#1#3}
\crefformat{subsubsection}{\S#2#1#3}
\crefrangeformat{section}{\S#3#1#4 to~\S#5#2#6}
\crefrangeformat{subsection}{\S#3#1#4 to~\S#5#2#6}
\crefrangeformat{subsubsection}{\S#3#1#4 to~\S#5#2#6}

\definecolor{darkblue}{rgb}{0.0, 0.0, 0.55}
\begin{document}

\author{Linus Gustafsson}
\affiliation{%
    \institution{TU Berlin}
    \city{Berlin}
    \country{Germany}}
\email{lig@3s.tu-berlin.de}
\orcid{0009-0001-7557-8742}

\author{Trever Schirmer}
\affiliation{%
    \institution{TU Berlin}
    \city{Berlin}
    \country{Germany}}
\email{ts@3s.tu-berlin.de}
\orcid{0000-0001-9277-3032}

\author{David Bermbach}
\affiliation{%
    \institution{TU Berlin}
    \city{Berlin}
    \country{Germany}}
\email{db@3s.tu-berlin.de}
\orcid{0000-0002-7524-3256}

\newcommand{\shortName}{Tremulant}
\title[\shortName]{\shortName: Latency in Serverless Platforms\\beyond the 99\textsuperscript{th} Percentile}


\begin{abstract}
    Function-as-a-Service (FaaS) increases developer productivity by removing operational concerns such as managing hardware and software runtimes.
    Because function instances are started on demand, some requests must wait for a new instance, called a cold start.
    These are a common obstacle to more widespread adoption of FaaS, as they lead to unpredictable latency spikes that are not acceptable for some use cases.
    To solve this, many cold start optimizations have been implemented.

    However, comparatively little attention has been directed towards the slowest \emph{warm} starts.
    With improvements in cold start times, slow warm starts emerge as a new problem for extreme tail events.
    By analyzing a dataset of several million requests across seven FaaS platforms, we show that the long tail of slow \emph{warm} starts can be hundreds of times slower than their median performance, while a virtual machine we use as a baseline is substantially more consistent.
    We confirm this across four independent public datasets spanning 2020 to 2025, so the effect is neither new nor specific to one platform.
    Moreover, the warm share among the slowest calls starts to rise again at the highest percentiles: the most extreme latencies are not only cold starts, and cold-start mitigation alone cannot remove them.
\end{abstract}

\maketitle

\newcommand{\TF}[1]{\ensuremath{\mathrm{TF}_{#1}}}
\newcommand{\WS}[1]{\ensuremath{\mathrm{WS}_{#1}}}

\section{Introduction}
\label{sec:introduction}

\begin{figure}
    \centering
    \includegraphics[width=\linewidth]{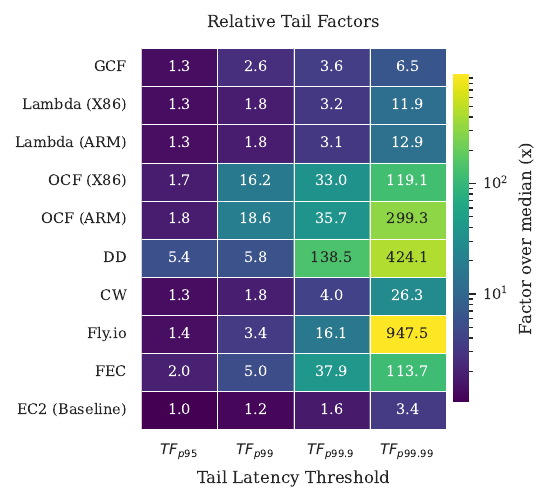}
    \caption{
    Even when only considering \emph{warm} starts, the highest latency percentiles are much slower than their median.
    In our baseline, AWS EC2, the p99 latency is less than \qty{20}{\percent} slower than its median.
    Even at p95, \emph{every} FaaS platform is further from its own median than EC2 at p99.
    At p99.99, warm starts are between \num{6.5} and \num{947.5} times slower than their median.
    }
    \label{fig:heatmap}
\end{figure}

Function-as-a-Service (FaaS) is a serverless cloud-computing model in which applications are composed of small, event-driven, ideally stateless functions, while operational responsibilities such as provisioning, scaling, and runtime management are handled by the cloud provider.
Since FaaS entered mainstream technology stacks, popularized by Amazon Web Services (AWS) Lambda, it has seen significant adoption due to its promise of simplicity: developers can focus primarily on application logic, which can improve iteration speed and productivity \cite{eismann_review_2021}.
This momentum has prompted other cloud providers to introduce their own FaaS offerings~\cite{hendrickson_serverless,ekwe-ekwe_state_2024}.
To extend FaaS to a broader set of latency-sensitive applications, research and engineering efforts increasingly focus on improving latency behavior and, in particular, latency consistency.
This is motivated both by user-facing quality requirements and by economics: because execution is commonly billed per invocation and execution duration, latency improvements can directly reduce cost.
In practice, two effects contribute strongly to inconsistency.
First, requests that require a new function instance to be started before processing (so-called cold starts) can substantially increase end-to-end latency \cite{bardsley_serverless_2018}.
Cold starts have been studied extensively, and many techniques exist to mitigate their impact in a wide range of scenarios \cite{manner_cold_2018,bermbach2020coldstarts}.
Very slow \emph{warm} invocations, however, have received less attention in research, even though they \emph{alone} already exhibit a long latency tail.
While cold starts are represented more in the higher latency percentiles than warm starts, they are far rarer than warm starts, and their share drops again at the highest latency percentiles.
In \cref{fig:heatmap}, we show the slowdown of warm starts at specific percentiles, compared to the platform's own median.
While the p99.99 request latency on our baseline virtual machine is \tFactor{3.4} slower than the median request, it is hundreds of times slower on multiple FaaS platforms. 
Complex serverless applications often comprise multi-function workflows~\cite{eismann_review_2021}, where the probability that at least one tail event occurs increases with every additional function call.

In this paper, we identify tail events of warm starts as a cross-platform issue and validate the effect against existing datasets.
For this, we present a reusable method of making heterogeneous serverless traces tail-comparable. %
We then validate our findings by analyzing existing datasets from other serverless research, demonstrating that these issues have existed for a long time and are not unique to one platform or dataset.
We publish all analysis available as open-source software.\footnote{\url{https://github.com/3s-rg-codes/warm-tail-analysis}}
In summary, we make the following contributions:

\begin{enumerate}
    \item We characterize warm-start latency across seven FaaS platforms and an IaaS baseline. We show that warm invocations can be more than two orders of magnitude slower than their median in the tail.
    \item We introduce the warm share as a metric to describe the composition of the slowest calls, and show that warm calls increasingly contribute to the extreme tail.
    \item We screen eleven public datasets and reproduce these effects in the four that qualify. 
\end{enumerate}

\section{Background}
\label{sec:background}
FaaS resides at a high abstraction level in the cloud computing spectrum~\cite{wang_2018_peeking}.
It abstracts away infrastructure management, runtime provisioning, and scheduling, so users only provide code as a deployment package that the platform loads into a function instance.
Although platforms implement it differently, all of them use virtualization technologies to isolate instances from each other~\cite{schirmer2026newKids}.
Some newer, edge-oriented platforms use the mechanism provided by the language runtime to isolate functions, e.g., the isolates provided by the JavaScript runtime V8~\cite{cf2020workers}.
Other platforms use lower-level mechanisms, e.g., containers~\cite{k8sImages} or microVMs~\cite{agache2020Firecracker}.
The isolation mechanisms influence the overhead of a cold start and the resource configurability options a user is provided with.

\paragraph{Function Lifecycle}
The cost-efficiency of FaaS is enabled by an execution model in which a function instance is started only when a request arrives.
At a high level, this model results in two common invocation paths that behave similarly across many FaaS platforms.
First, if no function instance is available to serve an incoming request (either because no instance exists or all are busy handling requests), the platform must create and initialize a new instance.
This process is commonly referred to as a \emph{cold start}~\cite{manner_cold_2018}.
Depending on the underlying virtualization technology, cold starts can consume substantial compute resources and add disproportionate latency because the platform must create a virtual environment and initialize the language runtime and dependencies~\cite{liu_faaslight_2023,ebrahimi_cold_2024}.
Therefore, cold starts are frequently associated with tail-latency events and can substantially degrade platform performance~\cite{bermbach2020coldstarts}.
Moreover, because billing is often tied to metered execution duration and initialization of the environment is considered part of the execution, cold starts can substantially increase cost~\cite{ebrahimi_cold_2024}.

Second, if an instance is available, the platform can route the request to it, avoiding the overhead of a cold start.
Such invocations are commonly referred to as \emph{warm starts}~\cite{wang_2018_peeking}.
To increase the likelihood of warm starts, many FaaS platforms keep instances alive for some time after request processing instead of terminating them immediately~\cite{schall_lukewarm_2022}.
Platforms terminate instances when demand subsides to save cost and resources, as users are typically not billed for this keep-alive period.

As increasingly complex use cases are implemented on FaaS platforms, applications often consist of multiple functions with explicit dependencies~\cite{eismann_review_2021}.
Such multi-function workflows are frequently coordinated through event-based processing and orchestration~\cite{bermbach2020coldstarts}.

\section{Related Work}
\label{sec:relatedwork}

FaaS platforms have attracted substantial research interest across benchmarking methodology, cold-start mitigation, and tail latency analysis; yet, as we discuss below, warm-start behavior at extreme latency percentiles remains largely unexplored.

One major area of interest has been understanding the proprietary architectures and technologies underlying commercial FaaS platforms.
In particular, several efforts introduce tools that simplify benchmarking by automating deployment and experimental workflows.
A prominent subcategory of these tools focuses on exposing FaaS platforms to realistic workloads to assess the applicability of the paradigm to business use cases.
For example,~\citet{grambow2021befaas,kim_functionbench_2019,yu_characterizing_2020} propose such frameworks.
Conversely, \citet{copik_sebs_2021} contribute a suite of automated microbenchmarks that characterize isolated aspects of FaaS platforms, such as cold-start behavior and latency variability, alongside more complex workloads for assessing cost efficiency and invocation overhead.

Although these frameworks primarily target end-to-end analysis and general platform performance, complementary work addresses more specific capabilities.
\citet{scheuner2026traceIt} focus on tracing by integrating their benchmarking framework with platform-specific tracing tools.
\citet{kuhlenkamp2020datasetISE} investigate platform elasticity by generating diverse load patterns across phases and correlating these phases with distinct performance modes across four platforms.
In practice, most benchmarking frameworks cover only a small subset of FaaS platforms; the most frequently studied are Google Cloud Functions, AWS Lambda, and Azure~\cite{scheuner_function-as--service_2020}.
In particular, modern edge FaaS platforms such as Cloudflare Workers (CW)~\cite{cf2020workers}, Deno Deploy (DD)~\cite{denoland2021deploy}, and Fastly Edge Compute (FEC)~\cite{fastly2019edge} remain underrepresented in research~\cite{schirmer2026newKids,scheuner_function-as--service_2020}.

Other work has focused specifically on minimizing the impact of cold starts on overall platform performance.
Two primary strategies have been pursued: (i) reducing the duration of individual cold starts~\cite{liu_faaslight_2023,du_catalyzer_2020}, and (ii) reducing their frequency~\cite{wang_replayable_2019}.
For the former, both academic and industry research has sought to lower the overhead of virtualization technologies while preserving the strong function isolation required for security guarantees~\cite{ao2022faasnap,akkus_SAND_2018,oakes_SOCK_2018}.
A notable contribution is Firecracker, developed by AWS and subsequently open-sourced by \citet{agache2020Firecracker}.
For the latter, researchers have explored the inherent dependencies between functions and proposed optimization strategies such as merging multiple functions into a single deployment~\cite{schirmer2022fusionize} or predicting function invocations to pre-warm instances before requests arrive~\cite{bermbach2020coldstarts}.

By contrast, warm-start tail latency has received comparatively little attention.
This gap reflects a prevalent assumption in the literature: that long tails in FaaS platforms are largely attributable to cold starts.
While this assumption has motivated extensive cold-start research, it has left the tail behavior of the warm path, the dominant operating mode in production deployments \cite{joosen2023datasetHuawei,joosen2025datasetHuawei,schirmer2026newKids}, underexplored.
Moreover, tail latency is commonly characterized only at the 99th percentile, leaving the remaining \qty{1}{\percent} of invocations unexamined.
We argue that this portion has a significant impact on observed performance, even when cold starts are excluded.
For instance, although \citet{ustiugov_analyzing_2021} analyze FaaS platforms with a focus on tail latencies, providing insight into their causes and effects, they only consider data up to the 99th percentile.

The study of extreme tail latencies has received more attention in adjacent fields, such as IaaS deployments and OS-level performance analysis.
In these contexts, latency beyond the 99th percentile is sometimes considered, yielding a more complete picture of how tail shape influences system-level performance~\cite{li2014tales}.
\citet{dean_tail_2013} show that even rare high-latency events in the top \qty{1}{\percent} of requests can have a significant cumulative performance impact when considered at scale, e.g., in large datacenters.
\citet{xu2013avoiding} examine the causes and effects of tail latencies in IaaS environments, finding that virtualization exacerbates long-tail behavior by factors of two to four.
\citet{hao_mittos_2017} address the problem at the OS level, showing how I/O scheduling can be adapted to limit the impact of tail latency events on application-level performance.

\section{Analyzing Tail Latency Events}
\label{sec:methodology}

\newcommand{\memlist}[2]{\qtylist[list-units = single, list-separator = {/}, list-final-separator = {/}]{#1}{#2}}
\begin{table}[t]
    \caption{
        The platforms and configurations present in the primary dataset by~\citet{schirmer2026newKids}.
        \emph{Reg.}\ (\cmark) marks regional platforms, where users pick a deployment region; \xmark{} marks worldwide platforms, which replicate deployments and route requests to a nearby location.
        The last two columns give the number of warm calls at or beyond p99.99.
        \footnotesize{
            $^\dagger$Regional configuration is also possible.
            $^\ddagger$A second version was released in 2024.
            $^\mathsection$FEC reports every invocation as a cold start (see \cref{sec:methodology}), so its count covers all calls.
        }
    }
    \label{tab:platformsIncluded}
    \footnotesize
    \setlength{\tabcolsep}{3pt}
    \begin{tabularx}{\linewidth}{@{}>{\raggedright\arraybackslash}Xclrr@{}}
        \toprule
        & & & \multicolumn{2}{c}{\bf{Warm calls at p99.99}} \\
        \cmidrule(l){4-5}
        \bf{Platform} & \bf{Reg.} & \bf{Memory} & \bf{ARM} & \bf{x86} \\
        \midrule
        AWS Lambda~\cite{aws2014lambda}                                  & \cmark & \memlist{128;512;1024}{\mega\byte} & \num{192} & \num{162} \\
        Cloudflare Workers (CW)~\cite{cf2020workers}                     & \xmark & not configurable & \multicolumn{2}{c}{\num{28}} \\
        Deno Deploy (DD)~\cite{denoland2021deploy}                       & \xmark & not configurable & \multicolumn{2}{c}{\num{29}} \\
        Fastly Edge Compute (FEC)~\cite{fastly2019edge}                  & \xmark & not configurable & \multicolumn{2}{c}{\num{64}$^\mathsection$} \\
        Fly.io$^\dagger$~\cite{flyio2020fly}                             & \xmark & \memlist{256;512;1024}{\mebi\byte} & -- & \num{187} \\
        Google Cloud Run Functions (GCF)$^\ddagger$~\cite{google2017gcf} & \cmark & \memlist{128;512;1024}{\mebi\byte} & -- & \num{192} \\
        Oracle Cloud Infrastructure Functions (OCF)~\cite{oracle2019ocf} & \cmark & \memlist{128;512;1024}{\mega\byte} & \num{189} & \num{185} \\
        \bottomrule
    \end{tabularx}
\end{table}

In this paper, we analyze an existing dataset comprising millions of calls to seven different FaaS platforms~\cite{schirmer2026newKids}.
The seven platforms are commercial FaaS platforms that are well documented, are not built on top of another FaaS platform, and do not use the default Kubernetes scheduler, which has been shown to have limited performance in FaaS implementations~\cite{Cvetković_2024_Dirigent}.
\Cref{tab:platformsIncluded} shows all platforms, their abbreviation used in this paper, and the respective function resource configurations present in the dataset.
While previous work based on this dataset focused on understanding the architecture of the platforms and its impact on overall performance, this paper focuses on warm tail latency events as a subset of the gathered data.

\paragraph{Execution}
In the primary dataset by~\citet{schirmer2026newKids}, platform performance is measured by repeatedly executing a minimal FaaS function on each platform.
The functions are empty functions that immediately return.
For each request (i) the function start time, and (ii) whether the request was served by a cold instance are recorded.
To detect a cold start, the functions use a global variable that is written during the first request.
The experiment is executed over an extended period to reduce bias from diurnal, weekly, and seasonal workload shifts that are common on FaaS platforms~\cite{schirmer2023nightshift}.
Moreover, the dataset’s extended experimental duration\footnote{The experiments by \citet{schirmer2026newKids} were conducted over 53 days, 2025-11-13 to 2026-01-05.} makes it particularly suitable for analyzing tail latency events:
The long timeframe yields a sufficiently large sample size, which is essential for reliably capturing and studying such inherently rare occurrences.
Additionally, due to the minimalist function setup that excludes outside influences, no tail events can be caused by user code, leaving the platform and network path as the only possible sources.
Functions are called with an inter-arrival time of \qty{3}{\second} for \qty{1}{\minute}, repeated every \qty{5}{\minute}.
Because requests arrive continuously within each iteration, and iterations are close together, this setup increases the likelihood of observing warm starts.

\paragraph{IaaS Baseline}
To contextualize tail latencies observed in the dataset, we adopt the approach of Copik et al.~\cite{copik_sebs_2021} and run the same warm start workload on a virtual machine (VM).
In contrast to FaaS, a request to an IaaS deployment does not require a platform-managed control plane or router.
Consequently, the request path is substantially shorter, avoiding potential sources of additional latency. 
In our experiments, the VM thus serves as a reference point for the level of consistency that a minimal IaaS instance can achieve.
We use the smallest possible AWS EC2 instance type (\texttt{t3.nano}), start it in the same region that the functions were deployed in, serve the same code from a Node.js HTTP server, and put it under the same load as the functions for \num{11} days.\footnote{The IaaS baseline experiments were conducted from 2026-04-16 to 2026-04-27.}
Although \texttt{t3.nano} instances have burstable CPU performance, we do not expect CPU credit exhaustion or resource contention to affect the measured latencies, since the request pattern, as in the primary dataset~\cite{schirmer2026newKids}, imposes only a very low load.
While the baseline experiment ran after the data collection period of the primary dataset, we argue that the effect size we identify is too large to be caused by seasonal differences.

\begin{figure*}
    \centering
    \includegraphics[width=\linewidth]{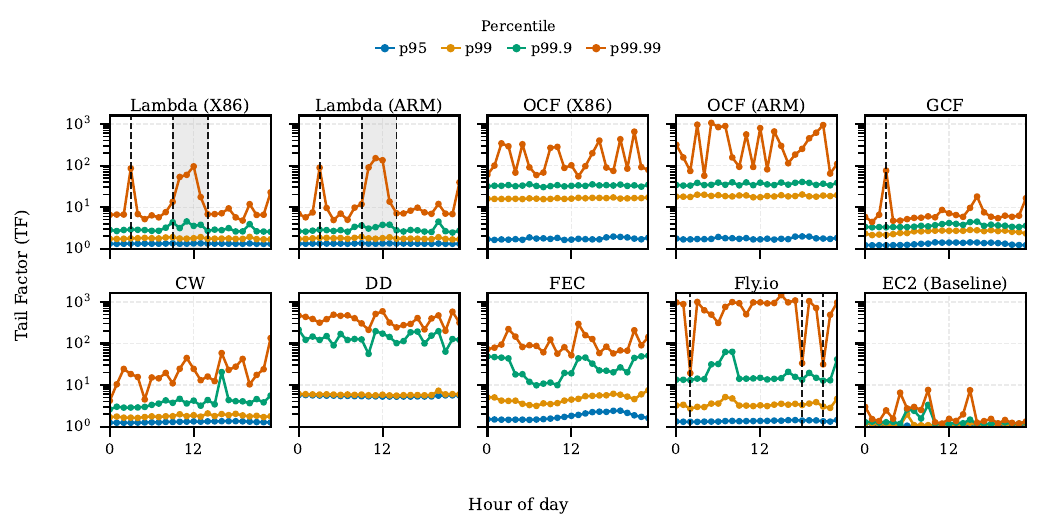}
    \caption{The tail factor $\TF{p^*}$ for p95, p99, p99.9, and p99.99 across all platforms.
    Generally, there is a pronounced divergence between p99.99 and the other evaluated percentiles, underscoring the importance of examining these extreme latencies to fully understand platform performance.
    OCF, FEC, Fly.io, CW, and DD demonstrate inconsistent tail performance throughout the day, whereas Lambda and GCF exhibit more stable behavior disrupted by isolated spikes, as indicated by the vertical lines.
    Furthermore, AWS EC2 and GCF show substantially lower relative tail latencies compared to the other FaaS platforms.
    Notably, the EC2 baseline exhibits the lowest overall relative tail latencies and the highest consistency.
    }
    \label{fig:daily-patterns}
\end{figure*}

\paragraph{Metrics}
\label{para:metrics}
Our analysis of tail latency relies on three core metrics:
The first metric is the latency percentile.
Although the 95\textsuperscript{th} (p95) and 99\textsuperscript{th} (p99) percentiles are the standard measures for characterizing tail behavior, higher percentiles have received comparatively little attention in the context of FaaS platforms.
To characterize the extreme tail in our measurements, we additionally report p99.9 and p99.99, following the \enquote{9s}-pattern.
Latency is measured as Time-To-First-Byte (TTFB), defined as the elapsed time until the client receives the first byte of the server response.
Rather than anchoring tail events to a single fixed percentile, we examine how the full range of percentiles collectively characterizes platform performance.

The second metric is the tail factor ($\TF{p^*}$), defined as the ratio of an arbitrary latency percentile $p^*$ to the median (p50):
$$
\TF{p^*} = \frac{\text{latency}_{p^*}}{\text{latency}_{p50}}
$$
We use this metric as we are largely interested in tail performance in relation to the median performance of a platform, as it makes the platforms comparable. 
This gives an intuitive understanding of how much worse the tail is compared to the expected median behavior.
$\TF{p^*}$ and latency percentiles are always calculated using only warm starts.
The only exception is FEC, which reports every invocation as a cold start because global state does not persist between its invocations; for FEC, we therefore use all calls.

The third metric is the warm share ($\WS{p^*}$).
Unlike the previous two metrics, it describes the \emph{composition} of the slowest calls.
For a latency percentile, it reports what fraction of requests in this percentile were warm starts:
$$
\WS{p^*} = \frac{n^{\text{warm}}_{\ge p^*}}{n_{\ge p^*}}
$$
A $\WS{}$ above \qty{50}{\percent} means that the majority of the slowest requests at or above this percentile were warm starts.
The absolute values can only compare composition within platforms, as the platforms have different base rates of cold requests.
To make them comparable, we additionally report the overall rate of cold starts per dataset.

\section{Tail Latency Across Seven Platforms}
\label{sec:experiments}

\begin{figure*}
    \centering
    \includegraphics[width=\linewidth]{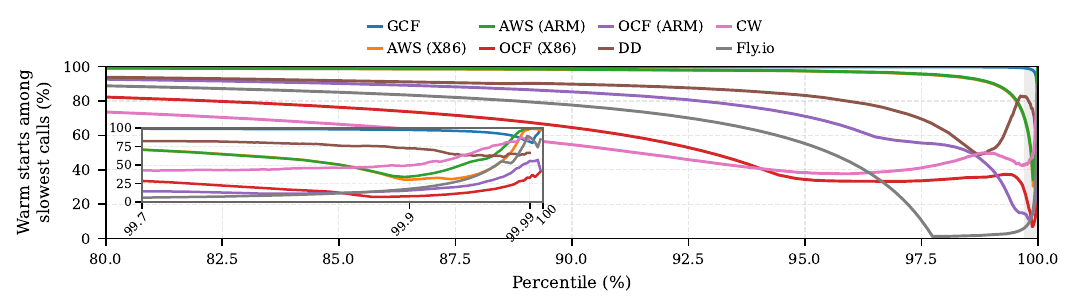}
    \caption{
        Warm start percentage of all calls in a latency percentile.
        At p80, most calls are warm starts for every provider shown.
        As expected, the rate generally drops in the higher percentiles, as cold starts usually take longer.
        The warm share then starts rising beyond its minimum on every platform except for GCF, which stays comparatively close to \qty{100}{\percent} throughout. 
        At p99.99, more than \qty{50}{\percent} of calls are warm starts on five out of six platforms.
        The effect is very clear on DD: $\WS{}$ is \qty{>90}{\percent} at p80, reaches its minimum of \qty{\sim 50}{\percent} around p99, and rises to \qty{\sim 75}{\percent} at p99.5 before easing off again in the extreme tail.
        This shows that the extreme tail does not just belong to cold starts.
        The per-provider lines end when \num{<30} data points remain.
        FEC is excluded from this figure, since all calls above p80 are cold starts.
    }
    \label{fig:warm-share}
\end{figure*}

We find that warm start tail latency behavior decreases platform performance substantially.
Calls in percentiles that are rarely considered when evaluating FaaS platforms can be up to \tFactor{947.5} slower than the median.
Moreover, for most platforms the highest latencies are not just caused by cold starts but also by warm starts.

\paragraph{Tail Severity Across Platforms}

Tail latency often worsens non-linearly, and a few extreme outliers (which are often disregarded) dominate the very end of the distribution.
In the dataset, we observe a recurring pattern, which is visualized in \cref{fig:heatmap}.
Latencies in p95 and p99 are often similar, even though the interval between them spans \qty{4}{\percent} of calls, expressing a denser latency distribution in that range.
However, the tail then steepens, with a jump at p99.9 (\qty{0.9}{\percent} of calls) and another large jump at p99.99 (\qty{0.09}{\percent}).
Across every platform, as well as on the IaaS baseline, the step from p99.9 to p99.99 is the largest single \emph{absolute} increase in $\TF{}$.

This pattern holds for all platforms except FEC and OCF, which exhibit wide and relatively evenly spaced latency distributions even above the 95th percentile.
For instance, DD exhibits a $\TF{p95}$ of \num{5.43} and a $\TF{p99}$ of \num{5.85}, but strongly increased $\TF{p99.9}$ and $\TF{p99.99}$ values of \num{138.46} and \num{424.05}, respectively.
That is, p95 and p99 are closely clustered, but p99.9 and p99.99 are more spread out and show a strong increase.
Consequently, the true extent of the tail and its impact often only become visible when observing values above p99.

This characterizes a long, heavy tail, which is most visible beyond the 99th percentile.
Such tails have a disproportionate impact on overall platform performance. 
For instance, when disregarding values above the 99th percentile for DD, we observe improvements of the distribution's mean of up to \qty{33}{\percent} in comparison to the mean of the full distribution.
Moreover, when disregarding the 99.99th percentile for Fly.io, removing just \qty{0.01}{\percent}, the mean improves by \qty{13}{\percent}.

\paragraph{Comparison to the IaaS Baseline}

\begin{figure}
    \centering 
    \includegraphics[width=\linewidth]{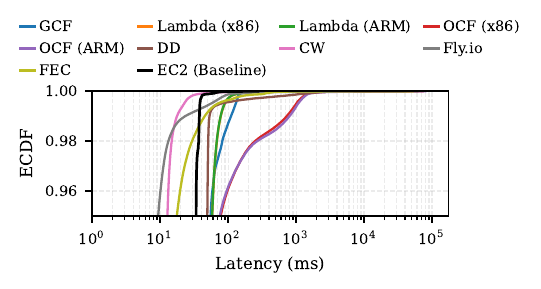}
    \caption{
    ECDF showing \emph{absolute} latency values above the 95th percentile for all providers.
    While CW, Fly.io and FEC have requests that complete faster than the baseline VM running on AWS EC2, it has a denser distribution and hence more consistent performance than all FaaS platforms with a $\TF{p99.99}$ of \num{3.4}.
    The FaaS platforms lie between \num{6.5} and \num{947.5}.
   }
    \label{fig:ecdf-absolute}
\end{figure}

Although the severity of tail distributions varies across observed FaaS platforms, all of them exhibit heavy tails.
In particular, \cref{fig:ecdf-absolute} visualizes that absolute latency values of FaaS platforms span one order of magnitude at p95, but at p99.9 they already span close to two orders of magnitude, with latencies reaching up to approximately \qty{1500}{\milli\second} for OCF. %
Contrarily, the latency distribution of the \textit{IaaS baseline} (AWS EC2 \texttt{t3.nano}) is substantially denser with tail latencies that are only marginally higher than the median.
Even above the 99.5th percentile, where all observed FaaS platforms experience strong performance degradation, latencies only slightly increase for the EC2 instance (see \cref{fig:ecdf-absolute,fig:heatmap}).
For instance, the $\TF{p99.99}$ for the AWS EC2 instance is \num{3.4}, while it is between \num{6.5} and \num{947.5} for the FaaS platforms.
In absolute values that means p99.99 is $\approx$ \qty{100}{\milli\second} for the IaaS baseline and $\approx$ \qty{6514}{\milli\second} for Fly.io as the worst performing FaaS platform in that regard.
Notably, the fastest requests from FEC, Fly.io, and CW outperform the IaaS baseline.
However, these platforms ultimately exhibit inferior performance at higher percentiles.
This early performance advantage is likely attributable to their deployment close to the benchmarking client, decreasing networking overhead.

\paragraph{Diurnal Variation at p99.99}
Although \cref{fig:daily-patterns} shows that p95, p99, and p99.9 differ considerably across platforms, their percentiles are largely stable diurnally within each platform.
However, p99.99 exhibits substantial diurnal variance.
For instance, GCF, Fly.io, and Lambda exhibit isolated but substantial latency spikes or dips at p99.99, while lower percentiles remain largely unaffected.
For OCF, CW, FEC, and DD, p99.99 (and partially p99.9) also shows unstable behavior, but events are less isolated and generally span longer periods of instability.
Although the IaaS baseline also shows isolated p99.99 spikes, their magnitude is less pronounced and hence aligns with the overall observed performance.
Particularly, \cref{fig:daily-patterns} visualizes that the p99.99 latency spikes for the IaaS baseline are up to two orders of magnitude lower than those of OCF, DD, and Fly.io.

Beyond the 99.99th percentile, an examination of the absolute maximum latency per platform provides further insight into the upper bounds of invocation delays.
However, since an absolute maximum represents a single isolated outlier, it lacks the statistical robustness of the p99.99 metric, which aggregates between \num{28} and \num{192} calls depending on the platform and architecture (\cref{tab:platformsIncluded}), and should be interpreted as a rare worst-case event rather than a stable performance indicator.

Fly.io exhibits the highest relative delay with a $\TF{p100}$ of approximately \num{6149}, corresponding to an absolute latency of \qty{42}{\second}.
DD and CW follow with $\TF{p100}$ values of roughly \num{4240} and \num{712}, respectively; FEC is omitted, as it has no warm starts.
Interestingly, GCF shows the lowest $\TF{p100}$ at \num{631}, an absolute maximum latency of approximately \qty{27}{\second}.

\paragraph{Warm Share}
\cref{fig:warm-share} shows what share of a latency percentile comprises warm starts.
Because the platforms cold-start at very different rates, these shares are only comparable across platforms once that base rate is known:
over the whole experiment, \qty{0.01}{\percent} of GCF calls are cold, \qty{0.2}{\percent} on both AWS configurations, \qty{1.2}{\percent} on DD, \qty{1.5}{\percent} on OCF ARM, \qty{2.2}{\percent} on Fly.io, \qty{3.6}{\percent} on OCF x86 and \qty{6.0}{\percent} on CW, while FEC reports every call as cold.
While the warm share consistently drops above p80, it starts rising again on almost all platforms.
For DD, it starts rising again just above p99, but then drops off in the extreme tail.
For all other platforms, the warm share starts increasing again eventually.
Even for GCF, where almost all calls are warm, the warm share starts increasing after p99.99.
At p99.9, six out of nine configurations are dominated by cold starts, i.e., $\WS{} \leq \qty{50}{\percent}$.
Past p99.99, the tail is dominated by warm starts for all platforms except OCF x86 and FEC, where virtually every call is a cold start.
While the lines in \cref{fig:warm-share} stop when fewer than \num{30} data points remain, the number of observations differs per platform.
At p99.9, they range from \num{297} on DD and CW to \num{1919} on GCF and AWS ARM.
At p99.95, they range from \num{148} to \num{960}.

\section{Validation with Existing Datasets}
\label{sec:validation}
\newcommand{\awsMahmoudi}{Lambda20$^{\alpha}$}
\newcommand{\awsKuhlenkamp}{Lambda20$^{\beta}$}
\newcommand{\azureKuhlenkamp}{Azure20}
\newcommand{\gcfKuhlenkamp}{GCF20}
\newcommand{\ibmKuhlenkamp}{IBM20}
\newcommand{\gcfSchirmer}{GCF23}
\newcommand{\huaweiJoosen}{Huawei25}

\providecommand{\yes}{\textcolor{OliveGreen}{\textbf{Yes}}}
\providecommand{\no}{\textcolor{RedOrange}{\textbf{No}}}
\begin{table*}
    \caption{Datasets used to validate the findings of the primary dataset, and the tail factor and warm share observed in each.
        We include a dataset if its per-request records contain both the request latency and whether the request was a cold start.
        \huaweiJoosen{} does not annotate cold starts explicitly, but it can be inferred from the podID.
        We screened and excluded seven additional datasets, which do not provide this information~\cite{shahrad2020dataseAzure2019,zhang2021datasetAzure2021,nasiri2026datasetIBM2026,joosen2023datasetHuawei,maissen2020datasetFaaSDOM,wang2021datasetFaaSNet,ginzburg2021datasetOSAArchive}.
        We count a dataset as showing a large tail factor if relative increases between percentiles are exponential or $\TF{p95} > 20$, and as showing a warm-share rebound if $\WS{}$ rises again after its minimum above p80, indicating a higher relative occurrence of slow warm starts.
        We refer to the datasets by the short forms defined in this table.\\
        \footnotesize{
            $^\dagger$\azureKuhlenkamp{} has a request failure rate of \qty{>95}{\percent}, limiting validity for Azure.
            $^\ddagger$These datasets use Google Cloud Functions, while the primary dataset uses the successor Google Cloud Run functions.
            $^\S$Their functions are computationally complex, so we exclude execution duration from our analysis.
        }
    }
    \label{tab:validation}
    \begin{tabularx}{\linewidth}{l l l X X}
        \toprule
        \bf{Platform}       & \bf{Dataset}                                              & \bf{Abbreviation}   & \bf{Tail Factor}                                              & \bf{Warm-Share Rebound}\\
        \midrule
        AWS Lambda          & \citet{kuhlenkamp2020datasetISE}                          & Lambda20$^{\beta}$  & \yes, much higher than in the primary dataset.                & \yes, from \qty{35.9}{\percent} at p99.05 to \qty{87.5}{\percent} at p99.9 \\
                            & \citet{mahmoudi2020datasetModelingServerlessPerformance}  & Lambda20$^{\alpha}$ & \yes, higher than in the primary dataset.                     & \yes, slightly, from \qty{83}{\percent} at p99.9 to \qty{98}{\percent} at p99.95 \\\addlinespace[1pt]
        Azure$^\dagger$     & \citet{kuhlenkamp2020datasetISE}                          & Azure20             & \no$^\dagger$                                                 & \no$^\dagger$ \\\addlinespace[1pt]
        GCF$^\ddagger$      & \citet{kuhlenkamp2020datasetISE}                          & GCF20               & \yes, much higher than in the primary dataset.                & \yes, slightly, from \qty{90}{\percent} at p99.75 to \qty{97.8}{\percent} at p99.95 \\
                            & \citet{schirmer2023nightshift}                            & GCF23               & \yes, higher than in the primary dataset.                     & \no, \qty{0.0}{\percent} warm from p80 to p99.9 \\\addlinespace[1pt]
        Huawei              & \citet{joosen2025datasetHuawei}                           & Huawei25$^\S$       & \yes / \no, strongly depends on resource configuration.       & \no, cold starts are at most \qty{5.7}{\percent} of the tail \\\addlinespace[1pt]
        IBM Cloud Functions & \citet{kuhlenkamp2020datasetISE}                          & IBM20               & \no, but lower relative factor caused by high median latency. & \yes, from \qty{46.6}{\percent} at p99.7 to \qty{66}{\percent} at p99.95 \\
        \bottomrule
    \end{tabularx}
\end{table*}

\begin{figure}[t]
    \centering
    \includegraphics[width=\linewidth]{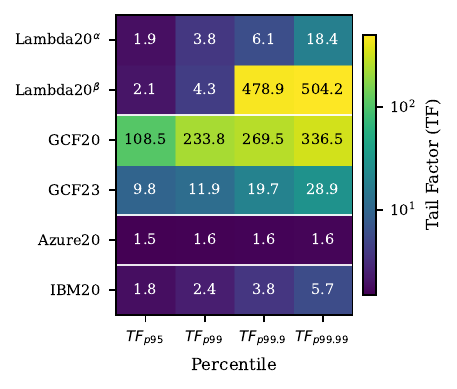}
    \caption{$\TF{}$ for all platforms included in the validation datasets. 
    Overall, the results demonstrate that tail effects extend beyond the platforms analyzed in the primary dataset. 
    Furthermore, comparing \gcfSchirmer{} and \gcfKuhlenkamp{} to the primary dataset reveals substantial increases, which are likely attributable to differences in platform versions. 
    Lambda is the only platform that also appears in the primary dataset. 
    \awsKuhlenkamp{} and \awsMahmoudi{} use the same platform.
    Their differing $\TF{}$ values show that the effect varies depending on load patterns. 
    The Azure results should be interpreted with caution, as \citet{kuhlenkamp2020datasetISE} report an error rate exceeding \qty{95}{\percent}.
    }
    \label{fig:heatmap-related-work}
\end{figure}

As a next step, we analyze existing research datasets for the effects we have observed, focusing on (i) the tail factor, (ii) the warm share, and (iii) diurnal variations.
This serves two purposes: 
First, it rules out that these effects are an artifact of the data or testing methodology.
Second, it improves the understanding of which platforms exhibit these effects and for how long they have existed.

\subsection{Methodology}
We make heterogeneous traces \emph{tail-comparable} in four steps.
We show the resulting included and excluded datasets in \cref{tab:validation}.
To be included, a dataset needs to contain per-request latency values and indicate whether a request is cold or warm.
Because the experiments were designed to evaluate different aspects of FaaS platforms, many experiments do not use empty functions like in the primary dataset.
As differing code execution time would invalidate comparability of results, the datasets must contain information that can be used to isolate platform performance.
For instance, client request latency (RTT) as well as code execution time are measured for most datasets, which allows subtraction of the two to isolate platform overhead.
This makes the datasets tail-comparable.
An exception to this is the dataset of \citet{joosen2025datasetHuawei}, which records \textit{frontend RTT} instead of the client RTT used by all other datasets.
Due to this, the tail-comparability of \citet{joosen2025datasetHuawei} is limited.
\huaweiJoosen{} is the only included dataset that does not annotate cold starts explicitly; there, we treat the first request on a given \texttt{podID} as cold and every later request on that pod as warm.
Finally, where a dataset ships two disagreeing encodings, we use the platform-reported one rather than the derived one: \texttt{CNew} for \citet{kuhlenkamp2020datasetISE} and \texttt{isCold} for \gcfSchirmer{}.

As mentioned in \cref{sec:methodology}, $\WS{}$ is influenced not only by the severity of cold starts but also by their frequency, and hence strongly depends on load patterns.
In the existing datasets, the base cold start rate varies from \qty{<0.1}{\percent} to \qty{>50}{\percent}.

\begin{figure*}
    \centering
    \includegraphics[width=\linewidth]{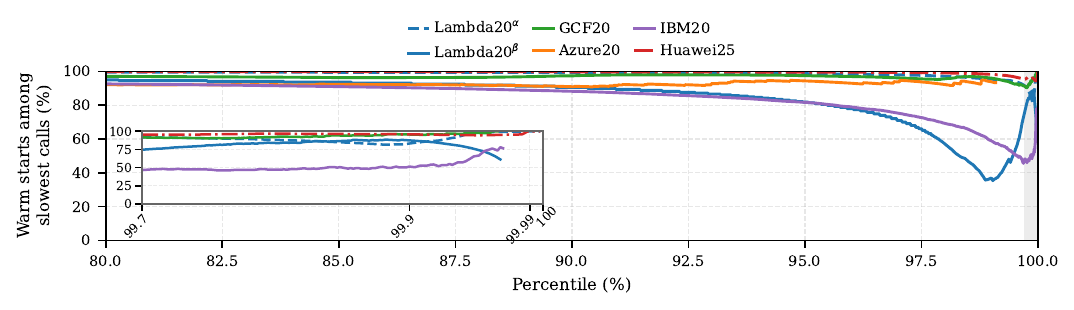}
    \caption{
    Warm share $\WS{}$ of latency percentiles in the validation datasets, analogous to \cref{fig:warm-share}.
    This figure excludes \gcfSchirmer{}, where all calls above p80 are cold.
    Otherwise, most datasets almost completely comprise warm starts.
    For the datasets that have a meaningful cold start percentage, the $\WS{}$ starts to increase at p99.05 for \awsKuhlenkamp{} and p99.7 for \ibmKuhlenkamp{}.
    }
    \label{fig:warm-share-related-work}
\end{figure*}

\subsection{Results}

The analysis of existing datasets provides strong evidence that the effects identified in the primary dataset are not unique. 
While their magnitude varies across platforms and datasets, tail-latency amplification and warm-tail behavior can be observed consistently throughout the analyzed literature (\cref{tab:validation}).

\begin{figure}
    \centering
    \includegraphics[width=\linewidth]{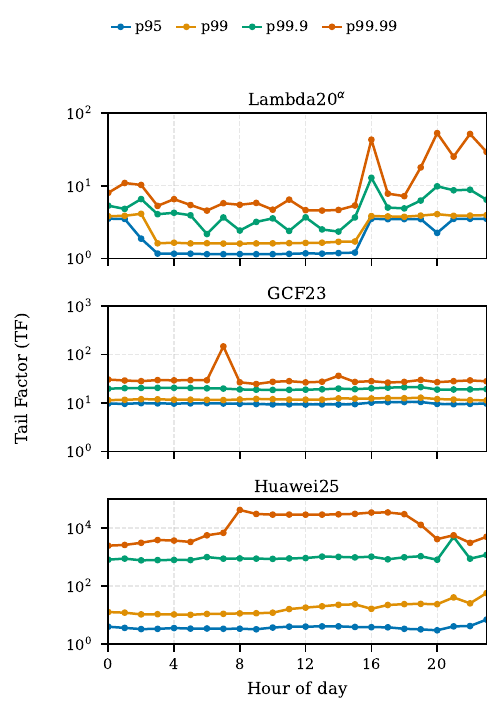}
    \caption{
        Tail factor $\TF{p^*}$ over the course of a day, analogous to \cref{fig:daily-patterns}.
        Performance consistency varies strongly across platforms, with $\TF{p99.99}$ peaking at \num{53} for \awsMahmoudi{}, \num{146} for \gcfSchirmer{}, and \num{41317} for \huaweiJoosen{}.
        In particular, \gcfSchirmer{} shows consistent, although increased latencies while the primary dataset and \awsMahmoudi{} show generally low latencies.
        \huaweiJoosen{} only contains relative timestamps; hence, their time is relative to the experiment start.
        \huaweiJoosen{} and \awsMahmoudi{} stem from a single day, increasing the probability of higher variance or the dominance of performance by an isolated event.
        The datasets from \citet{kuhlenkamp2020datasetISE} could not be analyzed for diurnal performance as the experiments did not run for long enough.
    }
    \label{fig:intra-day-related-work}
\end{figure}

\paragraph{AWS Lambda}
Lambda appears in the primary dataset as well as in \awsMahmoudi{} and \awsKuhlenkamp{} both from 2020 and using x86 processors.
The results show strong variations across datasets.
\awsMahmoudi{} indicates similar performance for AWS as in the primary dataset, except for a stronger increase of $\TF{p99.99}$ in the former (see \cref{fig:heatmap,fig:heatmap-related-work}).
\awsKuhlenkamp{} shows a median of \qty{29}{\milli\second} with a p99.99 of \qty{14623}{\milli\second}, while in \awsMahmoudi{} the median is \qty{92}{\milli\second} with a p99.99 of \qty{1703}{\milli\second}, resulting in a strong difference in $\TF{p99.99}$ for the two datasets.

This can be attributed to a difference in the respective experiment setups:
While in \awsMahmoudi{} the platform is exposed to a relatively constant load, it is deliberately tested for elasticity in \awsKuhlenkamp{}, leading to differences in cold start frequency.
For diurnal variability, \awsMahmoudi{} shows similar effects to those in the primary dataset.
That is, although the platform is generally stable with tightly clustered tail percentiles, it shows isolated spikes at 4~pm, 8~pm, and 10~pm (cf.~\cref{fig:intra-day-related-work}).
Additionally, p95 and p99 show consistent, bimodal performance with an increase during the night.

The warm share falls to \qty{35.9}{\percent} warm at p99.05 in \awsKuhlenkamp{}, the lowest minimum of any validation dataset shown in \cref{fig:warm-share-related-work}.
It then climbs back up to \qty{87.5}{\percent} at p99.9 (\num{96} calls).
Over the whole dataset, \qty{1.63}{\percent} of calls are cold.
In \awsMahmoudi{}, which has a lower overall cold rate of \qty{0.13}{\percent}, the $\WS{}$ is \qty{>95}{\percent} until p99, then decreases to \qty{83}{\percent}, and then increases to \qty{\sim 98}{\percent} at p99.95.

\paragraph{Azure Functions}
\azureKuhlenkamp{} shows the data for the benchmark by \citet{kuhlenkamp2020datasetISE} in 2020.
They note that \qty{> 95}{\percent} of their requests failed, limiting the usefulness of analyzing tail events.
Their warm share, shown in \cref{fig:warm-share-related-work}, is \qtyrange{91}{100}{\percent} throughout p80---p100, but the sample size is very small: \num{32} calls at p99 and \num{4} at p99.9.

\paragraph{Google Cloud Functions}
Google Cloud Functions v1 was analyzed in \gcfSchirmer{} in 2023 and \gcfKuhlenkamp{} in 2020.
It is a predecessor to Google Cloud Run Functions (also referred to as Google Cloud Functions v2), which is used in the primary dataset.
Both \gcfKuhlenkamp{} and \gcfSchirmer{} show that the older version is less stable, indicated by strong increases in all calculated $\TF{}$ values compared to the primary dataset, as visualized in \cref{fig:heatmap-related-work}.
Distinctly, \gcfSchirmer{} shows less pronounced $\TF{}$ values than \gcfKuhlenkamp{}, which exhibits a $\TF{p99}$ that is already above \num{100}.
That is, v1 shows p99.99 latencies as high as \qty{28000}{\milli\second} while v2 shows a p99.99 of only \qty{275}{\milli\second}.

The warm share for \gcfKuhlenkamp{} is high throughout the experiment, dropping to \qty{\sim 90}{\percent} at p99.75 before rising to \qty{>97}{\percent}.
\gcfSchirmer{}, however, has different behavior: the $\WS{}$ is \qty{0.0}{\percent} warm from p80 to p99.9, a clear separation of cold and warm starts.
The datasets have very different baseline rates: \qty{1.2}{\percent} for \gcfKuhlenkamp{} and \qty{52.7}{\percent} for \gcfSchirmer{}, which can be attributed to differences in workloads.

Diurnal variability is consistent in \gcfSchirmer{}, with the exception for an isolated spike at 7am.
Notably, the relative increases over the median are higher overall than in the primary dataset.
Due to the strong diurnal consistency, the hourly $\TF{}$ values shown in \cref{fig:intra-day-related-work} can also be seen in \cref{fig:heatmap-related-work}.
Overall, the results for Google Cloud Functions v1 are not as relevant for future research, as it has since been replaced.

\begin{figure}[t]
    \centering
    \includegraphics[width=\linewidth]{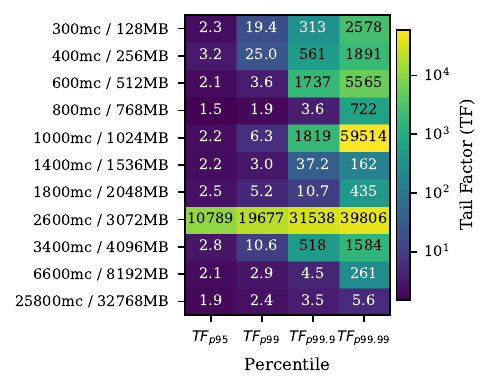}
    \caption{The equivalent of \cref{fig:heatmap}, detailing the $\TF{}$ for the various function configurations evaluated in Huawei25.
    CPU resources are expressed in millicores (mc) and memory in MB.
    Configurations with fewer than \num{1000} calls are omitted.
    Platform performance varies substantially across configurations, with the most popular setup, by number of calls, (\qty{1000}{\millicore} / \qty{1024}{\mega\byte}) exhibiting the highest $\TF{p99.99}$.
    This indicates that, in contrast to the other analyzed platforms, function configuration can also play a crucial role in shaping the distribution tail.
    Some resource configurations have a very high median, leading to lower $\TF{}$ even though absolute values are comparable to other configurations (cf.~\cref{par:huawei}).
    }
    \label{fig:huawei-heatmap}
\end{figure}

\paragraph{Huawei FunctionGraph}
\label{par:huawei}
Huawei uses its own serverless platform running on top of Kubernetes~\cite{chen2024yuanRong}.
As \huaweiJoosen{} is taken from a production environment in 2025, performance across functions and function configurations varies strongly (see \cref{fig:huawei-heatmap}).

\huaweiJoosen{} is the only dataset that exhibits substantial differences between configurations that are non-negligible in the tail percentiles.
In particular, the most popular resource size (\qty{1000}{\millicore} / \qty{1024}{\mega\byte}) has a $\TF{p99.99}$ of \num{59514}, which maps to an absolute \emph{platform} delay of \qty{62.1}{\second} with a median of only \qty{1.0}{\milli\second}.\footnote{Notably, latency is measured as frontend RTT and not client RTT and hence does not contain network delays.}
The \qty{2600}{\millicore} / \qty{3072}{\mega\byte} configuration is more extreme still: on a comparable median of \qty{2.7}{\milli\second}, degradation sets in far earlier, with a $\TF{p95}$ of already \num{10789} and a $\TF{p99.99}$ of \num{39806}, i.e.\ an absolute platform delay of \qty{108.9}{\second}.
At the other extreme, the \qty[group-digits=none]{25800}{\millicore} / \qty[group-digits=none]{32768}{\mega\byte} configuration has the highest median of all configurations (\qty{47.3}{\milli\second}) and by far the flattest tail, with a $\TF{p99.99}$ of only \num{5.6}.

Cold starts make up \qty{0.37}{\percent} of calls at p95, which rises to \qty{5.6}{\percent} at p99.95.
This is an over-representation by a factor of \tFactor{266} relative to their overall rate of \qty{0.021}{\percent}, but at the same time only a small part of the tail.

Moreover, diurnal variability is pronounced in \huaweiJoosen{} with a wide spread between observed percentiles. 
In particular, the 99.99th percentile spans more than one order of magnitude over the course of a day.
The percentiles are, however, not bound to each other: p95 and p99 peak late in the day (\num{7.0} and \num{57.3} at 11~pm), whereas p99.99 peaks at 8~am (\num{41317}) and p99.9 at 9~pm (\num{4887}).

\paragraph{IBM Cloud Functions}
IBM Cloud Functions was analyzed in \ibmKuhlenkamp{} and used Apache OpenWhisk as the underlying platform, but was deprecated in 2023~\cite{ibm2023openWhiskDeprecation}.
Although the low $\TF{}$ visualized in \cref{fig:heatmap-related-work} indicates stable platform performance, this reflects generally high latencies.
For instance, the $\TF{p99.99}$ of \num{5.6} maps to an absolute \emph{platform} latency of \qty{19.7}{\second}, which implies a median latency of \qty{3.51}{\second}.

The $\WS{}$ in \ibmKuhlenkamp{} declines from \qty{\sim 80}{\percent} at p95 to \qty{\sim 47}{\percent} at p99.7, indicating that cold starts are the majority of calls at this percentile.
At even higher percentiles, however, the warm share increases again: it is \qty{\sim 51}{\percent} at p99.9 (\num{99} calls) and \qty{66}{\percent} at p99.95 (\num{50} calls).
Again, the results are not as relevant for future research, as the platform is no longer available.
 
\paragraph{Summary}
Overall, we see a clear indication that the same observations from our dataset also exist elsewhere.
We found the same tail latency issues in older datasets from platforms that are in the primary dataset, and we see similar results for other platforms as well.
We argue that this makes it unlikely that the effect is an artifact of the primary benchmarking framework.
Thus, the long warm tail of serverless requests is a widespread issue for multiple platforms and has existed for a long time.

\section{Discussion \& Future Work}

In our evaluation, we find that FaaS platforms generally exhibit strong performance variability, as evidenced by heavy distribution tails.
As previous research has shown, this variability is partially caused by cold starts.
However, we show that warm starts also contribute to the extreme tail.
Although platforms exhibit this deterioration to varying degrees, all platforms perform considerably worse than a comparable IaaS deployment.
In particular, we observe this in the primary dataset and validate it through analysis of additional datasets from different years and across various platforms.

\paragraph{Awareness Gap}
The strongest relative performance deterioration occurs in the highest percentiles.
These ranges are frequently dismissed as insignificant outliers or unrepresentative of overall platform performance.

However, we find that this is a misconception, especially for FaaS platforms, since the very heavy upper distribution tails substantially influence overall performance.
Such performance impacts become even more significant when analyzing the most common applications of FaaS:
Eismann et al.~\cite{eismann_review_2021} find that the largest fraction of FaaS applications are web services, which means that users are directly affected by decreased performance.
Especially in online retail, this can affect profits, with a \qty{10}{\percent} increase in website loading times already leading to a \qty{4.2}{\percent} decrease in total sales~\cite{gallino_need_2023}.

\paragraph{Workflow Effect}
Furthermore, beyond the direct impact on end users, tail latencies introduce systemic performance issues due to the compositional nature of common FaaS workloads.
Because individual FaaS functions are typically small and contain little logic to improve reusability, applications frequently rely on workflows of multiple function invocations to fulfill a single client request~\cite{bermbach2020coldstarts}.
Consistent with this, \citet{eismann_review_2021} find that \qty{59}{\percent} of FaaS workflows contain \tRange{2}{10}{} functions, \qty{19}{\percent} contain \tRange{10}{1000}{} functions, and \qty{3}{\percent} contain \num{1000} functions or more.
Consequently, when an application requires many function calls to complete a task, the probability of experiencing at least one tail latency event naturally increases.
This amplification follows from the union bound, where repeated trials increase the chance of observing at least one rare event~\cite{dean_tail_2013}.
Under an independence assumption, if a single call exceeds a latency threshold with probability $p$, then a workflow with $n$ calls exceeds it at least once with probability $1-(1-p)^n$.
For small $p$, this probability grows approximately linearly with workflow length (i.e., $1-(1-p)^n \approx np$).
Because independence is not guaranteed in practice, this is an upper bound.

\paragraph{Warm-Start Tails as a New Problem}

For an individual request, being a cold start leads to higher latency:
In \huaweiJoosen{}, a cold start is \tFactor{266} more likely to be in p99.95 compared to all calls.
When cold starts are this rare, warm invocations still make up the majority of the extreme tail.
Even when observing only warm starts, p99.9 and p99.99 still introduce substantial deterioration to overall platform performance (\cref{fig:heatmap,fig:heatmap-related-work}).
Cold start mitigations cannot address these kinds of tail events at all.
Because the warm-start tail events we observed (i) follow diurnal patterns (see~\cref{fig:daily-patterns}) and (ii) occur even for empty functions, we attribute them to platform overhead, such as control-plane behavior and request routing.
For example, Lambda and GCF exhibit strong performance deterioration around 10--12~am and 3~am UTC respectively.
We hypothesize that this is caused by an increased platform-wide load, which leads not only to cold starts but also to warm requests slowing down.

\paragraph{Warm-Tail Impact Across Platforms}
Warm-tail impact differs markedly between platforms.
For Lambda, CW, and GCF, platforms that host a large share of the public web~\cite{wang_measuring_2024}, p95 to p99.9 are very consistent, and degradation occurs only occasionally at p99.99 or is generally low.
This relative stability likely reflects operational maturity at scale and makes these platforms more attractive from a tail-latency viewpoint.
Nevertheless, even the most stable FaaS platforms still underperform considerably when compared to a traditional IaaS deployment (see~\cref{fig:ecdf-absolute}).
Even if cold starts were theoretically eliminated (e.g., in a workload where functions are constantly in use), even the best among the analyzed FaaS platforms cannot compete with the consistency of a comparable IaaS deployment.

\paragraph{Applicability for Low-Latency Workloads}
However, FaaS should also be analyzed in the context of its use cases.
In particular, the paradigm is not designed to promise low-latency code execution and request processing.
Its design principles follow the promise of simplicity and low operational overhead on the user's side, as well as virtually infinite scalability \cite{eismann_review_2021}.
Although latency plays a significant role in the latter, and platforms work to improve elasticity (e.g., through new virtualization technologies), it is not the main focus.

Nevertheless, it is important for users and researchers to be aware of the impact that warm tail events can have, so as not to disregard warm starts as sources of substantial latency increases.

Additionally, it is important to highlight that our analysis is designed to examine tail events, which requires isolating them.
We did so by evaluating empty functions from the primary dataset or calculating platform delay in the additional datasets.
However, in real-world use cases, functions are not empty, and longer runtimes decrease the relative impact of both warm tail events and cold starts.
That is, users will experience lower relative tail latencies due to running actual workloads.
However, a recent serverless workload characterization reports that \qty{82}{\percent} of applications and \qty{96}{\percent} of invocations exhibit average execution times below one second~\cite{nasiri2026datasetIBM2026}.
Consequently, even modest warm-start tail overheads can account for a significant fraction of total execution time in many real-world deployments.

\paragraph{Mitigating Warm-Start Tails}
Future work could explore mitigation strategies that reduce the user-visible impact of rare warm-start tail events.
Mitigation is primarily the responsibility of platform operators, who can leverage occurrence patterns to identify and reduce platform variability.
On the client side, mitigation could be implemented by detecting unusually slow requests that are expected to be warm starts, e.g., because the function was invoked recently, and retrying them, which would avoid long waiting times for the longest tail.
The difficulty is in detecting whether a long in-flight request will be served by a warm or cold instance, since cancelling a cold start would lead to worse overall performance.
This approach does, however, not address the root cause.

\section{Conclusion}

This paper studies latency tails in FaaS with a focus on warm-start behavior.
We find that as cold-start impact decreases on modern offerings, warm-start tail events become a comparatively more prominent source of tail latency.
Accordingly, improving perceived FaaS latency consistency increasingly requires understanding and mitigating warm-start tail events, not only cold starts.

Moreover, we argue that percentiles beyond p99 should not be disregarded as unrepresentative outliers.
In particular, p99.9 and p99.99 often capture rare slowdowns, even when only observing warm starts, that dominate the extreme tail and can meaningfully affect end-to-end latency.
This matters especially for latency-critical and multi-function workflows, where many invocations are composed into a single user-visible request.
At the same time, these requirements may be fundamentally misaligned with the typical sweet spot of FaaS, which prioritizes operational convenience and elasticity over strict latency guarantees.
Evaluations of FaaS platforms should report p99.9 and p99.99 separately for their cold and warm invocations, and also report the sample count behind these values.

Even when completely ignoring cold starts and considering only warm invocations, FaaS platforms in our evaluation do not match the latency consistency of a comparable IaaS deployment.
Therefore, for workloads that require tight tail guarantees, IaaS-style deployments remain difficult to replace with FaaS, even under favorable conditions.

Finally, we observe similar tail-latency and warm-start effects in multiple existing datasets.
This suggests that our observations are not artifacts of a single dataset or testing methodology, but recurring properties that have existed across platforms and over time.

\begin{acks}
    Funded by the \grantsponsor{DFG}{Deutsche Forschungsgemeinschaft (DFG, German Research Foundation)}{https://www.dfg.de/en/} -- \grantnum{DFG}{495343202}.
    The authors want to thank Lucas Reinken, who supported this within the scope of a bachelor's thesis.
\end{acks}

\balance

\bibliographystyle{ACM-Reference-Format}
\bibliography{bibliography.bib}

\end{document}